\documentclass[11pt]{article}
\usepackage{amsfonts}
\usepackage{amsthm}
 \usepackage{bezier}

\usepackage{epic,eepic,amsmath,amssymb,color,graphicx}

\begin{document}

  \noindent
 {\Large{\bf {A maximally superintegrable  deformation of the $N$-dimensional\\[6pt] quantum Kepler--Coulomb system }}}

\medskip 
\medskip 
\medskip

\begin{center}
{\sc \'Angel Ballesteros$^a$, Alberto Enciso$^b$, Francisco J. Herranz$^{a}$,\\[4pt] Orlando Ragnisco$^{c,}$\footnote{
 Based on the contribution presented at ``XXIst International Conference on Integrable Systems and Quantum symmetries (ISQS-21)",
June 12--16, 2013,  
 Prague, Czech Republic} and Danilo Riglioni$^{d}$}
\end{center}
\medskip 
\medskip 

\noindent
$^a$ Departamento de F\'\i sica,  Universidad de Burgos,
E-09001 Burgos, Spain\\
{\tt{angelb@ubu.es,\quad fjherranz@ubu.es}}  \\[4pt] 
 $^b$ Instituto de Ciencias Matem\'aticas,  CSIC, E-28049 Madrid,
Spain\\
{\tt{ aenciso@icmat.es}} \\[4pt] 
 $^c$ Dipartimento di Fisica, Universit\`a di Roma Tre and Istituto Nazionale di Fisica Nucleare sezione di Roma Tre, Via Vasca Navale 84, I-00146 Roma, Italy\\
{\tt{ ragnisco@fis.uniroma3.it }} \\[4pt] 
 $^d$ Centre de Recherches Math\'ematiques, Universit\'e de Montr\'eal, H3T 1J4 2920 Chemin de la tour, Montreal, Canada\\
 {\tt{ riglioni@CRM.UMontreal.ca}}

  \medskip 
\bigskip
\bigskip

\begin{abstract} 
\noindent
The $N$-dimensional quantum Hamiltonian 
$$
\hat{H} = -\frac{\hbar^2 {|\mathbf{q}  }  | }{2(\eta  +| {\mathbf{q}} |)} {\mathbf{\nabla}}^2 - \frac{k}{\eta + |{\mathbf{q}} |}
$$

\noindent is shown to be exactly solvable for any real positive value of the parameter $\eta$.
Algebraically, this Hamiltonian system can be regarded as a new maximally superintegrable $\eta$-deformation of the $N$-dimensional Kepler--Coulomb Hamiltonian while, from a geometric viewpoint, this superintegrable Hamiltonian can be interpreted as a system on an $N$-dimensional Riemannian space with nonconstant curvature. The eigenvalues and eigenfunctions of the model are explicitly obtained, and the spectrum presents a hydrogen-like shape for positive values of the deformation parameter $\eta$ and of the coupling constant $k$.
\end{abstract} 

\vfil

\newpage



\section{Introduction}

Let us consider the $N$-dimensional ($N$D) classical Hamiltonian given by 
\begin{equation}
\label{hclassica}
\mathcal{H}_\eta ({\mathbf{q}}, {\mathbf{p}}) = \mathcal{T}_\eta ({\mathbf{q}}, {\mathbf{p}}) + \mathcal{U}_\eta({\mathbf{q}}) = \frac{|{\mathbf{q}}| {\mathbf{p}}^2}{2(\eta + |{\mathbf{q}}|)} - \frac{k}{\eta + |{\mathbf{q}}|} ,
\end{equation}
where $k$ and  $\eta$ are
   real parameters,       ${\mathbf{q}}=(q_1,\dots,q_N)$, ${\mathbf{p}}=(p_1,\dots,p_N)$ $\in \mathbb{R}^N$ are conjugate coordinates and momenta, and
   ${\mathbf{q}}^2\equiv  |{\mathbf{q}}|^2=\sum_{i=1}^Nq_i^2$. We recall that   $\mathcal{H}_\eta$ has been proven to be a maximally superintegrable Hamiltonian by making use of symmetry techniques~\cite{sigma72011}. This means that $\mathcal{H}_\eta $ is endowed with the maximum possible number of $(2N-1)$ functionally independent constants of motion (including $\mathcal{H}_\eta $ itself). 
   
   Explicitly, $(2N-3)$ of such  integrals   are  provided by the radial symmetry of the system, namely,
\begin{equation}
\mathcal{C}^{(m)} = \sum_{1 \leq i < j \leq m} (q_i p_j - q_j p_i)^2, \qquad \mathcal{C}_{(m)} = \sum_{N-m < i < j \leq N} (q_i p_j - q_j p_i)^2, \qquad m=2,\dots,N;
\label{xa}
\end{equation}
such that $\mathcal{C}^{(N)} =\mathcal{C}_{(N)} \equiv  {\mathbf L}^2$ is the square of the total angular momentum. Furthermore, $\mathcal{H}_\eta $ is endowed with an $N$D Laplace--Runge--Lenz vector ${\mathbf R}$. This means there exist $N$ additional constants of motion coming from the components of ${\mathbf R}$, which are given by
\begin{equation}
\mathcal{R}_i = \sum_{j=1}^N p_j (q_j p_i - q_i p_j) + \frac{q_i}{|{\mathbf{q}}|} (\eta \mathcal{H}_\eta + k),\qquad i=1,\dots, N.
\label{xb}
\end{equation}
The squared modulus of  ${\mathbf R}$ is radially symmetric, and turns out to be expressible in terms of $\mathcal{H}_\eta$ 
and ${\mathbf L}^2$:
$$
{\mathbf R}^2=  \sum_{i = 1}^N \mathcal{R}_i^2 = 2 {\mathbf L}^2 \mathcal{H}_\eta + (\eta \mathcal{H}_\eta + k)^2 .
$$
We remark that each of the three sets $\{ \mathcal{H}_\eta , \mathcal{C}^{(m)} \}$, $\{ \mathcal{H}_\eta , \mathcal{C}_{(m)} \}$ $(m=2,\dots,N)$ and  $ \{ \mathcal{R}_i \}$ $(i = 1,\dots,N)$
is formed by $N$ functionally independent functions in involution, and the set $\{ \mathcal{H}_\eta, \mathcal{C}^{(m)}, \mathcal{C}_{(m)}, \mathcal{R}_i \}$ for $m=2,\dots,N$ with a fixed $i$ provides the set of $(2N-1)$ functionally independent functions. As it was shown in \cite{sigma72011}, the set of constants of the motion $\mathcal{R}_i$ can be obtained explicitly by appying a St\"ackel transform~\cite{Stackel2,Stackel4} to the nondeformed Kepler--Coulomb (KC)  Hamiltonian.

  This maximally superintegrable Hamiltonian is obviously endowed with an $\mathfrak{so}(N)$ Lie--Poisson symmetry, since it can be constructed on an $N$D spherically symmetric  space. In particular, we can consider the $N(N-1)/2$ generators of rotations $J_{ij} = q_i p_j - q_j p_i$ with $i < j$ and $i,j = 1,\dots,N$ which span the $\mathfrak{so}(N)$ Lie--Poisson algebra with Poisson brackets given by
$$
\{J_{ij} , J_{ik} \} = J_{jk}, \qquad \{J_{ij} , J_{jk} \} = -J_{ik}, \qquad \{J_{ik} , J_{jk} \} = J_{ij}, \qquad i < j < k    .
$$
Hence 
  the $(2N-3)$ angular momentum integrals $\mathcal{C}^{(m)}$ and $\mathcal{C}_{(m)}$ (\ref{xa})  turn out to be the quadratic Casimirs of some rotation subalgebras $\mathfrak{so}(m) \subset \mathfrak{so}(N)$. Moreover we   observe that the `$\eta$-deformation' of the    Laplace--Runge--Lenz vector $\mathcal{R}_i$ (\ref {xb}) closes the same Poisson algebra as its nondeformed counterpart, namely:
$$
\{J_{ij} , \mathcal{R}_k \} =  \delta_{ik} \mathcal{R}_j - \delta_{jk} \mathcal{R}_i , \qquad
\{\mathcal{R}_i , \mathcal{R}_j \}  =  -2 \mathcal{H}_\eta J_{ij} .
$$
Therefore, since
$$
\{J_{ij} ,  \mathcal{H}_\eta \} =\{\mathcal{R}_i   ,  \mathcal{H}_\eta \}=0,\qquad \forall i,j,
$$
the Hamiltonian  $\mathcal{H}_\eta$ behaves as a `constant'  with respect to the    $N(N+1)/2$ `generators' $\{  J_{ij}, \mathcal{R}_i \}$.
  This fact makes possible to identify the classical integrals of the motion with the generators of an $\mathfrak{so} (N+1)$ algebra similarly to what happens with the usual  $N$D   Euclidean KC system (see~\cite{KC} and references therein).

We stress that maximally superintegrable Hamiltonians in $N$ dimensions are quite scarce,  even on the Euclidean space. The two representative examples of this class of systems are the KC system and the isotropic harmonic oscillator, for which all bounded trajectories are periodic (Bertrand's Theorem). In this respect, we recall that a maximally superintegrable `deformation/generalization' of the $N$D isotropic oscillator was firstly presented   
in~\cite{PD} and its quantum counterpart was constructed and solved in~\cite{darbouxiii, annals2011}.
 In fact, on the same footing of such a  maximally superintegrable oscillator system, the system $\mathcal{H}_\eta$ ({\ref{hclassica})  can be regarded as a genuine (maximally superintegrable) $\eta$-deformation of the $N$D usual KC system, since the limit $\eta \rightarrow 0$ of $\mathcal{H}_\eta$ (\ref{hclassica})  yields
$$
\mathcal{H}_0 = \frac{1}{2} {\mathbf p}^2 - \frac{k}{|{\mathbf{q}}|} .
$$   

Moreover, from a geometric perspective the kinetic energy term $\mathcal{T}_\eta$ can be interpreted as the one generating the geodesic motion of a particle with unit mass on a conformally flat space $\mathcal{M}^N= (\mathbb{R}^N,g)$, which is the complete Riemannian manifold with metric
\begin{equation}
\label{metricataubnut}
{\rm d}s^2 = \left(1 + \frac{\eta}{|{\mathbf q}|}\right) {\rm d} {\mathbf q}^2
\end{equation} 
and nonconstant scalar curvature given by
$$
R = \eta (N-1)\frac{4(N-3)r + 3(N-2) \eta}{4r (\eta + r)^3 },
$$
 where we have introduced the radial coordinate $r = |{\mathbf q}|$. Here it is straightforward to check that the limit $\eta \rightarrow 0$ of (\ref{metricataubnut})  returns the flat Euclidean metric ${\rm d}s^2 = {\rm d} {\mathbf q}^2$ with $R= 0$.
In fact, we stress that $\mathcal{H}_\eta$   can be  naturally related to the Taub-NUT system~\cite{Ma82,AH85,GM86,FH87,GR88,IK94,IK95,uwano,BCJ,BCJM,GW07,JL}  since   $\mathcal{M}^N$ can be regarded as  the (Riemannian)   $N$D Taub-NUT space~\cite{annals324}. 

The aim of this contribution is to anticipate the main results concerning a maximally superintegrable quantization of the classical Hamiltonian (\ref{hclassica}), since a complete study of this new exactly solvable quantum system will be given elsewhere~\cite{preparation}.   In the next Section,  the properties of the classical  system, including  its effective potential, are presented. In Section 3  the corresponding quantum Hamiltonian is constructed  by imposing the existence of the quantum analog of the full set of  $(2N-1)$  classical integrals of the motion (\ref{xa}) and (\ref{xb}).  Finally, the explicit solution of the spectral problem  is sketched in Section 4.


\section{The classical Hamiltonian and its effective potential}

Firstly, it should be   remarked that, quite surprisingly, the potential $\mathcal{U}_\eta$  (\ref{hclassica}) with $\eta \neq 0$ can be considered as the $N$D generalization of an `intrinsic' oscillator on the curved space $\mathcal{M}^N$. This statement comes from the approach introduced in \cite{annals324}, that generalizes the Bertrand's Theorem~\cite{Bertrand2} to 3D conformally flat Riemannian spaces, thus providing a more general notion of KC and harmonic oscillator potentials~\cite{Perlick, Bertrand,commun}. To be self-contained, let us briefly recall  these ideas by considering a 3D spherically symmetric   space ${\mathcal{M}^3}$ with coordinates $(q_1,q_2,q_3)$ and equipped with  a metric 
$$
g_{ij} = f(|{\bf q}|)^2  \delta_{ij} ,
$$ 
such that $f(|{\bf q}|)=f(r)$ is the conformal factor. 
Then the corresponding  Laplace--Beltrami operator on    $\mathcal{M}^3$   is given by
$$
\Delta_{\mathcal{M}^3}=\sum_{i,j=1}^3\frac1{\sqrt g} \frac{\partial}{ \partial {q_i} }\sqrt g\, g^{ij}\frac{\partial}{ \partial {q_j} },
\label{wa}
$$
where 
$g^{ij}$ is the   inverse  of the   metric tensor   $g_{ij}$, and $g$  is the determinant of $g_{ij}$.
  The radial symmetric Green function $U(|{\bf q}|)=U(r)$ on $\mathcal{M}^3$ (up to multiplicative and additive
constants) is defined as the positive nonconstant solution to the equation
$$
\Delta_{\mathcal{M}^3}U(r)= 0
\quad\text{on}\quad \mathcal{M}^3\backslash\{\textbf0\}\, ,
\label{wb}
$$
namely,
\begin{equation}
U(r) =\int^r\frac{{\rm d} r'}{r'^2f(r')} .
\label{wc}
\end{equation}
The prescription  for the $N$D case~\cite{PD}  is to keep the very same definitions given in~\cite{Bertrand,commun,EP06d,EP08, LT87, LT95}  for the KC and oscillator potentials on    $\mathcal{M}^3$. In particular, the
\emph{intrinsic KC potential}  on  the $N$D space $\mathcal{M}^N$ will be defined by
\begin{equation}
{\cal U}_{\rm KC}(r) :=A\,U (r)+B ,
\label{wd}
\end{equation}
while the
\emph{intrinsic   oscillator potential}  is defined to be proportional
to the inverse square of the KC potential
\begin{equation}
{\cal U}_{\rm O}(r) :=\frac C{U^2(r)} +D,
\label{we}
\end{equation}
where  $A,B,C$ and $D$ are     real constants.

Hence according to the above definitions and by 
considering the Taub-NUT  metric (\ref{metricataubnut}), with conformal factor
$$
f(r)=\sqrt{1+\frac{\eta}{r}} ,
$$
 it is straightforward to obtain  that the corresponding intrinsic potentials read
$$  
\mathcal{U}_{\rm KC} (r)   =A   \sqrt{1 + \frac{\eta}{r}}+B , \qquad 
\mathcal{U}_{\rm O} (r)=C  \frac{r}{r+ \eta }+D   .
$$
Therefore, whenever $\eta\ne 0$, we  find that  the  potential (\ref{hclassica}) corresponds  to  $\mathcal{U}_{\rm O}$ provided that $C=k/\eta$ and $D=-C$:
$$
\mathcal{U}_\eta (r)    = \frac{k}{\eta}\left(  \frac{ r }{r+ \eta }  -1  \right) =
 - \frac{k}{\eta + r},
 $$
which shows that $\mathcal{U}_\eta$ can be interpreted as an intrinsic oscillator on $\mathcal{M}^N$ with metric  (\ref{metricataubnut}).

In order to   understand the dynamical properties of the system (such as the existence of bounded states and any other critical features)  we shall make its radial symmetry manifest by introducing the hyperspherical coordinates $r, \theta_j$ $(j=1,\dots,N-1)$:
\begin{equation}
q_j = r \cos \theta_j \prod_{k=1}^{j-1} \sin \theta_j, \quad 1 \leq j < N, \qquad q_N = r \prod_{k=1}^{N-1} \sin \theta_k .
\label{xgf}
\end{equation} 
Their corresponding canonical momenta $p_r, p_{\theta_j}$  can be straightforwardly computed, and in these hyperspherical variables the $N$D Hamiltonian (\ref{hclassica})  takes the following 1D radial form 
\begin{equation}
\mathcal{H}_\eta (r, p_r) = \mathcal{T}_\eta(r, p_r) + \mathcal{U}_\eta(r) = \frac{r}{2(\eta + r)} \left( p_r^2 + \frac{{\mathbf L}^2}{r^2}\right) - \frac{k}{\eta + r} ,
\label{xf}
\end{equation} 
where the total angular momentum is given by 
$$
{\mathbf L}^2 = \sum_{j=1}^{N-1} p_{\theta_j}^2 \prod_{k=1}^{j-1} \frac{1}{\sin^2 \theta_k }.
$$


\begin{figure}
   \includegraphics[width=0.9\textwidth]{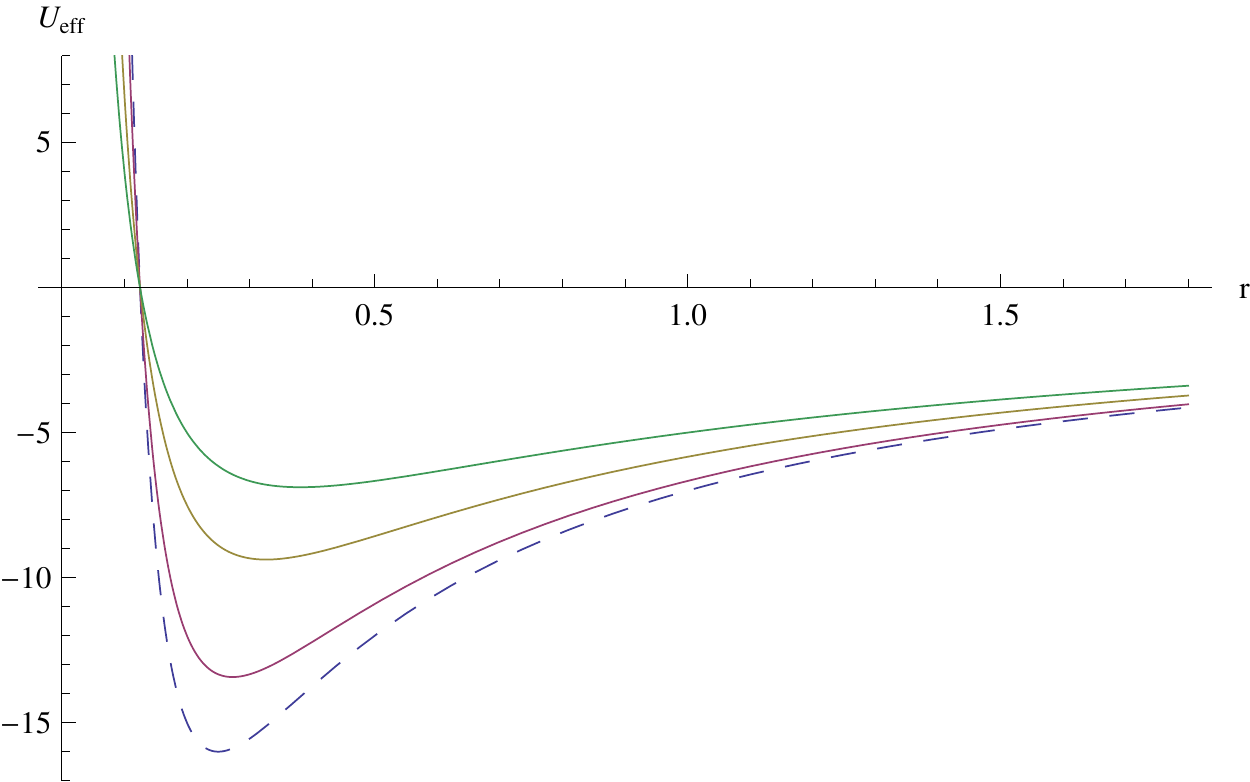}
\caption{The effective potential $ {\cal U}_{\rm eff} $ (\ref{effectiveham}) with $k=8$ and ${\mathbf L}^2=2$,  for $\eta=\{0,\, 0.05,\, 0.2,\, 0.4\}$. The dashed line corresponds to the KC potential ($\eta=0$)  and increasing values of $\eta$ lead to higher values of the potential.}
\label{figure2}        
\end{figure}


The   contribution to the dynamics of both of the non-flat metric and the potential can be better understood if we consider a   new set of canonical variables $Q,P$ given by
\begin{eqnarray}
&&Q (r)= \sqrt{r(\eta + r)} + \eta \log (\sqrt{r} + \sqrt{r+\eta}) , \nonumber\\
&& P(r,p_r)= \sqrt{\frac{r}{\eta + r}} \,p_r . \nonumber
\end{eqnarray}
In terms of these new variables the   Hamiltonian  (\ref{xf}) is  written as
$$
\mathcal{H}_\eta(Q,P) = \frac{1}{2}P^2  + \frac{{\mathbf L}^2}{2r(Q) (\eta + r(Q))}+ \mathcal{U}(r(Q)) \equiv  \frac{1}{2}P^2  + \mathcal{U}_{\rm eff}(Q),
$$
where the effective potential is thus given by
\begin{equation}
\label{effectiveham}
 {\cal U}_{\rm eff}(Q(r)  )= \frac{{\mathbf L}^2 }{2r(\eta+ r)} -\ \frac{k} {\eta+ r} .
\end{equation}
Consequently,   the radial motion can be described as the 1D dynamics of a particle under  the effective potential $ {\cal U}_{\rm eff}(Q (r))$.  As it can be appreciated from figure 1, 
     the radial equation admits a hydrogen-like potential for $\eta>0$ and $k>0$, which can be interpreted as a genuine $\eta$-deformation of the effective potential for the KC system.


\section{A maximally superintegrable quantization}

In order to obtain the quantum analog of the kinetic energy term $\mathcal{T}_\eta$  (\ref{hclassica}) we have to deal with the unavoidable ordering problems in the canonical quantization process that come from the nonzero curvature of the underlying space (see, e.g \cite{annals2011} and references therein). 
A detailed analysis of the different possible quantization prescriptions together with a proof of their equivalence through gauge transformations will be presented in a forthcoming paper \cite{preparation}.
One of this prescriptions consists in the so called `direct'  or Schr\"odinger quantization~\cite{uwano}, under which the quantum Hamiltonian $\mathcal{H}_\eta$ keeps the maximal superintegrability property and is therefore endowed with $(2N-1)$ algebraically independent operators that commute with $\mathcal{H}_\eta$.
This prescription has been already succesfully used in the case of a curved (Darboux III) oscillator system \cite{darbouxiii,annals2011} and makes use of all the algebraic machinery coming from the symmetries of the classical Hamiltonian. This result can be stated as follows.

\medskip

\noindent
{\bf Theorem 1.}  {\em Let $\hat{\mathcal{H}}_\eta$ be the quantum Hamiltonian given by

\begin{equation}
\label{htaubq}
\hat{\mathcal{H}}_\eta = \frac{|\hat{\mathbf q} |}{2 (\eta +|\hat {\mathbf q} |)}\, \hat{{\mathbf p}}^2 - \frac{k}{\eta +|\hat {\mathbf q}|} = \frac{|{\mathbf q}|}{2 (\eta + {|\mathbf q}|)} \left( -\hbar^2 \nabla^2 - \frac{2 k}{{|\mathbf q}|}\right) ,
\end{equation}
where $\hat{ \mathbf {q} }= \mathbf {q}$, $\hat{ \mathbf {p}}= -{\rm i}\hbar\nabla$ and $\nabla=(\partial_1,\dots,\partial_N)$ 
   such that $[ \hat q_i , \hat{p}_j] = i \hbar \delta_{ij}$. For any value of $\eta$ and $k$  it is verified that:}
\newline {\it (i) $\hat{\mathcal{H}}_\eta$ commutes with the following     operators ($ m= 2,\dots ,N; i=1,\dots,N$)
\begin{equation}
\hat{\mathcal{C}}^{(m)} = \sum_{1 \leq i < j \leq m} (\hat q_i \hat{p}_j -\hat q_j \hat{p}_i )^2 , \qquad \hat{\mathcal{C}}_{(m)} = \sum_{N-m \leq i < j \leq N} (\hat q_i \hat{p}_j - \hat q_j \hat{p}_i )^2 ,
\label{xh}
\end{equation}
$$
\hat{\mathcal{R}}_i = \frac{1}{2} \sum_{j=1}^{N} \hat{p}_j (\hat q_j \hat{p}_i - \hat q_i \hat{p}_j) + \frac{1}{2} \sum_{j=1}^{N}  (\hat q_j\hat{p}_i -\hat  q_i \hat{p}_j) \hat{p}_j + \frac{ \hat q_i}{\hat{{\mathbf q}}} \left(\eta \hat{\mathcal{H}}_\eta  +k  \right)     ,
$$ 
where $  \hat{\mathcal{C}}^{(N)} = \hat{\mathcal{C}}_{(N)}\equiv \hat{{\mathbf L}}^2$ is the total quantum angular momentum and  $$
\hat{\mathbf{R}}^2= \sum_{i=1}^N \hat{\mathcal{R}}_i^2 = 2 \hat{\mathcal{H}}_\eta \left(\hat{{\mathbf L}}^2 + \hbar^2 \frac{(N-1)^2}{4} \right) + \left(\eta \hat{\mathcal{H}}_\eta+k\right )^2 .
  $$
\newline  (ii) Each of the three sets $\{ \hat{\mathcal{H}}_\eta , \hat{\mathcal{C}}^{(m)} \}$, $\{\hat{\mathcal{H}}_\eta , \hat{\mathcal{C}}_{(m)} \}$ $(m= 2,\dots,N)$  and $\{ \hat{\mathcal{R}}_i\}$ $(i=1,\dots,N)$ is formed by $N$ algebraically independent commuting operators.
\newline  (iii) The set $\{ \hat{\mathcal{H}}_\eta , \hat{\mathcal{C}}^{(m)}, \hat{\mathcal{C}}_{(m)}, \hat{\mathcal{R}}_i \}$ for $m=2,\dots,N$ with a fixed index $i$ is formed by $2N-1$ algebraically independent operators 
\newline (iv) $\hat{\mathcal{H}}_\eta$ is formally self-adjoint on the Hilbert space $L^2( \mathcal{M}^N )$, endowed with the scalar product
$$
\langle \Psi | \Phi \rangle = \int_{\mathbb{R}^N} \overline{\Psi ({\mathbf q})} \Phi ({\mathbf q}) \left(1 + \frac{\eta}{{|\mathbf q}|} \right) {\rm d}{\mathbf q} .
$$
}
\medskip

The proof of this result can be obtained through direct computation. Recall that in \cite{annals2011} we have considered a similar  problem on the conformally flat Darboux III space.

Therefore,  the Hamiltonian (\ref{htaubq}) leads to the following Schr\"{o}dinger equation
$$
\left( \frac{-\hbar^2 {|\mathbf q}|}{2(\eta +| { \mathbf q } |)} \nabla^2 -  \frac{k}{\eta + {| \mathbf q }| }\right) \Psi ( {\mathbf q}) = E  \Psi ( {\mathbf q})   ,
$$
which in hyperspherical variables (\ref{xgf})  turns into
\begin{equation}
\label{htaubnutradialq}
\frac{  r}{2(\eta + r)} \left( -\hbar^2 \partial_r^2 - \frac{\hbar^2 (N-1)}{r} \partial_r + \frac{\hat{{\mathbf L}}^2}{r^2} - \frac{2 k}{r} \right) \Psi(r, \boldsymbol{\theta}) = E \Psi (r, \boldsymbol{\theta}) ,
\end{equation}
where $\boldsymbol{\theta} = (\theta_1,\dots, \theta_{N-1})$.

We remark that all the above results are well defined for any value of the parameters $\eta$ and $k$. Nevertheless, the explicit solution of the quantum Hamiltonian depends on the sign of both of them. In particular, hereafter we shall restrict ourselves to consider $\eta>0$, which implies that the variable $r\in(0,+\infty)$, and $k>0$ which corresponds to the proper `curved' hydrogen atom  potential.


\section{Spectrum and eigenfunctions}

In view of the effective potential introduced in (\ref{effectiveham}), one should expect that the quantum Hamiltonian (\ref{htaubq}) should have both a discrete and a continuous spectrum, and this is indeed the case.
When the Schr\"{o}dinger equation (\ref{htaubnutradialq}) is expressed in hyperspherical variables, it can be solved by factorizing the wave function into   radial and angular components 
$$
\Psi (r , \boldsymbol{\theta}) = \psi(r) Y(\boldsymbol{\theta}) , 
$$
and by considering the separability provided by the first integrals $\hat{  \mathcal{C} }_{(m)}$  (\ref{xh})  with eigenvalue equations given by
$$
  \hat{\mathcal{C}}_{(m)} \Psi = c_m \Psi, \qquad m=2,\dots,N .
$$
 From it, we obtain that $Y(\boldsymbol{\theta})$ solves completely the angular part and this corresponds, as expected,   to the hyperspherical harmonics  satisfying that
$$
  \hat{\mathcal{C}}_{(N)} Y(\boldsymbol{\theta})   \equiv \hat{{ \mathbf L }}^2 Y(\boldsymbol{\theta}) = \hbar^2 l (l + N -2) Y (\boldsymbol{\theta}), \quad l=0,1,2\dots
$$
where $l$ is the angular momentum quantum number. It can be proven~\cite{annals2011} that the eigenvalues $c_m$  are related with the $N-1$ quantum numbers of the angular observables through 
$$
c_k \leftrightarrow l_{k-1}, \qquad k=2,\dots,N-1, \qquad c_N \leftrightarrow l,
$$ 
which means that
$$
Y(\boldsymbol{\theta})\equiv Y^{c_N}_{c_{N-1},\dots,c_2}(\theta_1,\theta_2,...,\theta_{N-1}) \equiv Y^{l}_{l_{N-2},\dots,l_1}(\theta_1,\theta_2,...,\theta_{N-1}) .
$$
As a consequence, the radial Schr\"{o}dinger equation (\ref{htaubnutradialq})  is given by
$$
\frac{  r}{2(\eta + r)} \left( -\hbar^2\frac{{\rm d}^2}{{\rm d}r^2}  - \frac{\hbar^2 (N-1)}{r} \frac{{\rm d}}{{\rm d}r}  + \frac{\hbar^2 l(l+N-2)}{r^2} - \frac{2 k}{r} \right) \psi(r) = E  \psi (r) ,
$$
which can be written in the form
\begin{equation}
\label{radialcoupling}
 \left( -\hbar^2 \frac{{\rm d}^2}{{\rm d}r^2}   - \frac{\hbar^2 (N-1)}{r} \frac{{\rm d}}{{\rm d}r}  + \frac{\hbar^2 l(l+N-2)}{r^2} - \frac{2 K}{r} \right) \psi(r) =2 E  \psi (r) ,
\end{equation}
 where the new `coupling constant' $K$ turns out to be energy--dependent"
\begin{equation}
 K = k +  \eta E .
\label{za}
\end{equation}
 In this way we find that  the equation (\ref{radialcoupling}) is formally equivalent to the Schr\"{o}dinger equation of the radial hydrogen atom, whose bounded eigenfunctions are given, in terms of the generalised Laguerre polynomials $L^\alpha_n$, by:
\begin{equation}
\psi_{n,l} (r)    = r^l \exp\left( - \frac{K r}{\hbar^2\left (n+l+\frac{N-1}{2} \right)}\right) L_n^{2l+N-2} \left( \frac{2 K r}{\hbar^2\left(n+l+\frac{N-1}{2}\right)}\right) .
\label{eigenf}
\end{equation}
Notice that the eigenfunctions do not only depend   on the  usual quantum numbers $n,l$ but also on the eigenvalue $E$ through $K$ (\ref{za}).

If we substitute these functions within the equation  (\ref{radialcoupling}),  the following algebraic equation is obtained:
$$
E= - \frac{K^2}{2 \hbar^2 \left(n+l+\frac{N-1}{2}\right)^2} = - \frac{\left (k+ \eta  E\right)^2}{2 \hbar^2 \left(n+l+\frac{N-1}{2}\right)^2} .
$$
By solving such a quadratic equation in terms of $E$ we obtain the bounded discrete spectrum of the system, whose eigenvalues depend both on the `deformation' parameter $\eta$ and on the quantum numbers $n,l$. Namely,
\begin{equation}
E^\eta_{n,l} = \frac{-\hbar^2\left (n+l+\frac{N-1}{2}\right)^2 -\eta k + \sqrt{\hbar^4 \left(n+l+\frac{N-1}{2}\right)^4 +2 \eta k \hbar^2\left (n+l+\frac{N-1}{2}\right)^2}}{\eta^2} .
\label{spectrum}
\end{equation}
In this way,  the   eigenfunctions, $\psi_{n,l}^\eta (r)$,  
can be explicitly obtained by introducing (\ref{spectrum})  into $K$ (\ref{za}) and, next, by substituting the latter  in 
(\ref{eigenf}).
Note that in the limit $\eta\to 0$ the following well--known expression for the energies is recovered:
$$
E^0_{n,l} = -\frac{k^2}{2 \hbar^2 \left(n+l+\frac{N-1}{2}\right)^2} .
$$
And the first-order  effect of the deformation on the spectrum can be appreciated through a power series expansion in $\eta$:
\begin{equation}
\label{spettrolegato}
E^\eta_{n,l} =E^0_{n,l}  + \eta\,\frac{ k^3}{2 \hbar^4 \left(n+l+\frac{N-1}{2}\right)^4} + o(\eta^2) .
\end{equation}

 Let us remark that since the spectrum of $\hat{\mathcal{H}}_\eta$ is bounded from below (as the classical effective potential indicates), for a sufficiently large $k$ we can safely assume that $K = k + \eta E >0$. Moreover the condition $\Psi \in L^2 (\mathcal{M}^N)$ translates to 
$$
\int_{\mathbb{R}^N} |\Psi ({\mathbf q})|^2 \left( 1+ \frac{\eta}{|{\mathbf q}|}\right) {\rm d}{\mathbf q} < \infty .
$$
Note also  that the degeneracy of this spectrum is exactly the same as in the $N$D  hydrogen atom, which is a strong signature of the maximal  superintegrability of this quantum system. The explicit expressions for the wave functions corresponding to the continous spectrum will be given in the forthcoming paper \cite{preparation}.

  Finally let us stress that if we consider the following reparametrization 
  $$\eta \rightarrow \frac{1}{\sqrt{\lambda}} ,\qquad k \rightarrow \frac{\omega^2}{\sqrt{\lambda}} .
  $$
 the spectrum (\ref{spectrum}) turns out to have the same dependence on the principal quantum number (which in this case corresponds to $\mathcal{N} =n+l$) as the bounded spectrum for the intrinsic oscillator on a Darboux III space that was given in \cite{darbouxiii,annals2011}.     Indeed, this fact can be understood in terms of the classification of intrinsic KC and oscillator systems on non-Euclidean spaces provided in \cite{chino}. According to such classification, the Taub-NUT and the Darboux III oscillator are superintegrable systems of type II with the same parameters $\lambda,\delta$ (Taub-NUT: $\gamma = \frac{1}{2}, \lambda= 0, \delta$; Darboux III: $\gamma = 1, \lambda= 0, \delta$). However as showed in \cite{danilo} for the intrinsic KC systems, the parameter $\gamma$ only affects  the form of the principal quantum number $\mathcal{N}$ (Taub-NUT $\mathcal{N} = n+l$;  Darboux III: $\mathcal{N} = 2n+l$) but not the overall dependence on $\mathcal{N}$ which indeed turns out to be the same.


\section*{Acknowledgments}

This work was partially supported by the Spanish MINECO through the Ram\'on y Cajal program (A.E.) and under grants  MTM2010-18556 (A.B and F.J.H.), AIC-D-2011-0711 (MINECO-INFN)  (A.B, F.J.H.~and O.R.) and   FIS201-22566 (A.E.), by  the ICMAT Severo Ochoa under grant SEV-2011-0087 (A.E.),  by Banco Santander-UCM under grant GR35/10-A-910556 (A.E.), and by  a postdoctoral fellowship  by the Laboratory of
Mathematical Physics of the CRM, Universit\'e de Montr\'eal (D.R.).



\footnotesize


 \begin{thebibliography}{99}
 

\bibitem{sigma72011}  Ballesteros A, Enciso A,  Herranz F J,  Ragnisco O and    Riglioni D 2011 {\em  SIGMA} {\bf  7}  048



 \bibitem{Stackel2}
Kalnins E G, Kress J M and  
 Miller W Jr 2005
{\em J. Math. Phys.} {\bf 46}   053510 

  \bibitem{Stackel4}
Kalnins E G, Kress J M  and  
 Miller W Jr 2006
{\em J. Math. Phys.} {\bf 47}   043514 


\bibitem{KC} Ballesteros A, and  Herranz F J  2009  {\em J. Phys. A: Math. Theor.} {\bf  42} 245203


   

\bibitem{PD} Ballesteros A, Enciso A,  Herranz F J and Ragnisco O   2008  {\em Physica D} {\bf  237} 505

  

\bibitem{darbouxiii} Ballesteros A, Enciso A,  Herranz F J,  Ragnisco O and    Riglioni D 2011  {\em Phys. Lett. A} {\bf  375} 1431


\bibitem{annals2011}  Ballesteros A, Enciso A,  Herranz F J,  Ragnisco O and    Riglioni D   2011 {\em  Ann. Phys. } {\bf 326} 2053








  
 
\bibitem{Ma82}
 Manton N S 1982  {\em Phys. Lett. B} {\bf 110}  54 

\bibitem{AH85}
 Atiyah M F and   N.J.  Hitchin N J 1985  {\em Phys. Lett. A} {\bf 107}  21



\bibitem{GM86}
 Gibbons G W and    Manton N S  1986  {\em Nucl. Phys. B} {\bf 274} 183




\bibitem{FH87}
 Feh\'er L G and     Horv\'athy P A 1987  {\em Phys. Lett. B} {\bf 183}    182


\bibitem{GR88}
 Gibbons G W and   Ruback P J  1988  {\em Comm. Math. Phys.} {\bf 115}   267



\bibitem{IK94}
 Iwai T and  Katayama N   1994   {\em J.. Phys. A: Math. Gen.}  {\bf 27}  3179



\bibitem{IK95}
 Iwai T and  Katayama N 1995  {\em J. Math. Phys.} {\bf 36}   1790


\bibitem{uwano}
 Iwai T,   Uwano Y and   Katayama N 1996 {\em  J. Math. Phys.} {\bf 37}  608





 \bibitem{BCJ}  Bini D,  Cherubini C and Jantzen R T  2002  {\em Class. Quantum Grav.}  {\bf 19}    5481

\bibitem{BCJM} Bini D, Cherubini C, Jantzen R T and Mashhoon B 2003  {\em Class. Quantum Grav.}  {\bf {20}}  457

 \bibitem{GW07}
 Gibbons G W and  Warnick C M  2007   {\em J. Geom. Phys.} {\bf 57}  2286


\bibitem{JL}  Jezierski J and    Lukasik M   2007   {\em Class. Quantum Grav.}  {\bf 24}   1331

  
 



\bibitem{annals324} Ballesteros A, Enciso A,  Herranz F J and Ragnisco O 2009  {\em  Ann. Phys. } {\bf 324}  1219 


\bibitem{preparation}  Ballesteros A, Enciso A,  Herranz F J,  Ragnisco O and    Riglioni D  In preparation




\bibitem{Bertrand2}
  Bertrand J 1873  {\em C.R. Acad. Sci. Paris} {\bf  77 } 849


\bibitem{Perlick}
 Perlick V 1992  {\em Class. Quantum Grav.} {\bf  9}  1009

\bibitem{Bertrand}
Ballesteros A, Enciso A,  Herranz F J and Ragnisco O  2008   {\em Class. Quantum Grav.}  {\bf 25}    165005


\bibitem{commun}
Ballesteros A, Enciso A,  Herranz F J and Ragnisco O  2009  {\em Commun. Math. Phys.}   {\bf 290}     1033





\bibitem{EP06d}
 Enciso A and  Peralta-Salas D 2007 {\em J. Geom. Phys.} {\bf 57}  1679

\bibitem{EP08}
 Enciso A and  Peralta-Salas D  2009  {\em Indiana Univ. Math. J.}   {\bf 58} 1947
  
 


\bibitem{LT87}
 Li P and  Tam L F 1987   {\em Amer. J. Math.}  {\bf 109}  1129


\bibitem{LT95}
Li P and  Tam L F   1995 {\em J. Differential Geom.} {\bf 41} 277

 


\bibitem{chino} Ballesteros A, Enciso A,  Herranz F J,  Ragnisco O and    Riglioni D 2013
{\em Nankai Series in Pure, Appl. Math. Theor.  Phys.}  {\bf  11} ed C Bai, J P Gazeau and Mo-Lin Ge   (Singapore: World Scientific)    p 211, 
  arXiv:1304.4544

 



\bibitem{danilo} Riglioni D, {\em J. Phys. A: Math.   Theor.} {\bf 46}   265207

\end{thebibliography}
\end{document}